\documentstyle[psfig,11pt,emulateapj]{article}

\begin{document}
\title{The power spectral properties of the Z source GX~340+0}
\author{Peter G. Jonker\altaffilmark{1}, Michiel van der
Klis\altaffilmark{1}, Rudy
Wijnands\altaffilmark{1}$^{,}$\altaffilmark{2}, Jeroen
Homan\altaffilmark{1}, Jan van
Paradijs\altaffilmark{1}$^{,}$\altaffilmark{3}, Mariano
M\'endez\altaffilmark{1,4}, Eric C. Ford\altaffilmark{1}, Erik
Kuulkers\altaffilmark{5,6}, Frederick K. Lamb\altaffilmark{7}}

\altaffiltext{1}{Astronomical Institute ``Anton Pannekoek'',
University of Amsterdam, and Center for High-Energy Astrophysics,
Kruislaan 403, 1098 SJ Amsterdam; peterj@astro.uva.nl,
michiel@astro.uva.nl, homan@astro.uva.nl, mariano@astro.uva.nl,
ecford@astro.uva.nl} \altaffiltext{2}{MIT, Center
for Space Research, Cambridge, MA 02139, Chandra Fellow; rudy@space.mit.edu}
\altaffiltext{3}{University of Alabama, Huntsville}
\altaffiltext{4}{Facultad de Ciencias Astron\'omicas y
Geof\'{\i}sicas, Universidad Nacional de La Plata, Paseo del Bosque
S/N, 1900 La Plata, Argentina} \altaffiltext{5}{Space Research
Organization Netherlands, Sorbonnelaan 2, 3584 CA Utrecht, The
Netherlands; e.kuulkers@sron.nl} \altaffiltext{6}{Astronomical
Institute, Utrecht University, P.O. Box 80000, 3507 TA Utrecht, The
Netherlands} \altaffiltext{7}{Department of Physics and Astronomy,
University of Illinois at Urbana-Champaign, Urbana, IL 61801;
f-lamb@uiuc.edu}

\begin{abstract}
\noindent
We present an analysis of $\sim$390 ksec of data of the Z source
GX~340+0 taken during 24 observations with the {\em
Rossi\,X\,-ray\,Timing\,Explorer} satellite. We report the discovery
of a new broad component in the power spectra. The frequency of this
component varied between 9 and 14 Hz, and remained close to half that
of the horizontal branch quasi-periodic oscillations (HBO). Its rms
amplitude was consistent with being constant around $\sim$5\%, while
its FWHM increased with frequency from 7 to 18 Hz. If this sub-HBO
component is the fundamental frequency, then the HBO and its second
harmonic are the second and fourth harmonic component, while the third
harmonic was not detected. This is similar to what was recently found
for the black hole candidate XTE~J1550--564. The profiles of both the
horizontal- and the normal branch quasi-periodic oscillation peaks
were asymmetric when they were strongest. We describe this in terms of
a shoulder component at the high frequency side of the quasi-periodic
oscillation peak, whose rms amplitudes were approximately constant at
$\sim$4\% and $\sim$3\%, respectively. The peak separation between
the twin kHz quasi-periodic oscillations was consistent with being
constant at 339$\pm$8 Hz but a trend similar to that seen in,
e.g. Sco~X--1 could not be excluded. We discuss our results within the
framework of the various models which have been proposed for the kHz
QPOs and low frequency peaks.

\end{abstract}

\subjectheadings{accretion, accretion disks --- stars: individual
(GX\,340+0) --- stars: neutron --- X-rays: stars}

\section{Introduction}
\label{intro}
\noindent
GX\,340+0 is a bright low-mass X-ray binary (LMXB) and a Z source
(Hasinger \& van der Klis 1989).  The Z-shaped track traced out by Z
sources in the X-ray color-color diagram or hardness-intensity diagram
(HID) is divided into three branches: the horizontal branch (HB), the
normal branch (NB), and the flaring branch (FB). The power spectral
properties and the HID of GX\,340+0 were previously described by van
Paradijs et al. (1988) and Kuulkers \& van der Klis (1996) using data
obtained with the EXOSAT satellite, by Penninx et al. (1991) using
data obtained with the Ginga satellite, and by Jonker et al. (1998)
using data obtained with the {\em Rossi\,X\,-ray\,Timing\,Explorer}
{\em(RXTE)} satellite. An extra branch trailing the FB in the HID has
been described by Penninx et al. (1991) and Jonker et al. (1998).
When the source is on the HB or on the upper part of the NB,
quasi-periodic oscillations (QPOs) occur with frequencies varying from
20--50 Hz: the horizontal branch quasi-periodic oscillations or HBOs
(Penninx et al. 1991; Kuulkers \& van der Klis 1996; Jonker et
al. 1998). Second harmonics of these HBOs were detected by Kuulkers \&
van der Klis (1996) and Jonker et al. (1998) in the frequency range
73--76 Hz and 38--69 Hz, respectively. In the middle of the NB, van
Paradijs et al. (1988) found normal branch oscillations (NBOs) with a
frequency of 5.6 Hz.  Recently, Jonker et al. (1998) discovered twin
kHz QPOs in GX\,340+0. These QPOs have now been seen in all six
originally identified Z sources (Sco~X--1, van der Klis et al. 1996;
Cyg X--2, Wijnands et al 1998a; GX~17+2, Wijnands et al. 1997b;
GX~349+2, Zhang et al. 1998; GX~340+0, Jonker et al. 1998; GX~5--1,
Wijnands et al. 1998b; see van der Klis 1997, 1999 for reviews), but
not in Cir~X--1, which combines Z source and atoll source
characteristics (Oosterbroek et al. 1995; Shirey, Bradt, and Levine
1999; see also Psaltis, Belloni, \& van der Klis 1999).  \par
\noindent
In the other class of LMXBs, the atoll sources (Hasinger \& van der
Klis 1989), kHz QPOs are observed as well (see van der Klis 1997, 1999
for reviews).  Recently, also HBO-like features have been identified
in a number of atoll sources (4U\,1728--34, Strohmayer et al. 1996,
Ford \& van der Klis 1998, Di Salvo et al. 1999; GX13+1, Homan et
al. 1998; 4U~1735--44, Wijnands et al. 1998c; 4U~1705--44, Ford, van
der Klis, \& Kaaret 1998; 4U\,1915--05, Boirin et al. 1999;
4U~0614+09, van Straaten et al. 1999; see Psaltis, Belloni, \& van der
Klis 1999 for a summary). Furthermore, at the highest inferred mass
accretion rates, QPOs with frequencies near 6 Hz have been discovered
in the atoll sources 4U\,1820--30 (Wijnands, van der Klis, \&
Rijkhorst 1999c), and XTE~J1806--246 (Wijnands \& van der Klis 1998e,
1999b; Revnivtsev, Borozdin, \& Emelyanov 1999), which might have a
similar origin as the Z source NBOs.  \par
\noindent
At low mass accretion rates the power spectra of black hole
candidates, atoll, and Z sources show similar characteristics (van der
Klis 1994a,b).  Wijnands \& van der Klis (1999a) found that the break
frequency of the broken power law which describes the broad-band power
spectrum, correlates well with the frequency of peaked noise
components (and sometimes narrow QPO peaks) observed in atoll sources
(including the millisecond X-ray pulsar SAX\,J1808.4--3658; Wijnands
\& van der Klis 1998d, Chakrabarty \& Morgan 1998), and black hole
candidates.  The Z sources followed a slightly different correlation.
In a similar analysis, Psaltis, Belloni, \& van der Klis (1999) have
pointed out correlations between the frequencies of some of these QPOs
and other noise components in atoll sources, Z sources, and black hole
candidates, which suggests these phenomena may be closely related
across these various source types, or at least depend on a third
phenomenon in the same manner. Because of these correlations, models
describing the kHz QPOs which also predict QPOs or noise components in
the low-frequency part of the power spectrum can be tested by
investigating this low-frequency part.  \par
\noindent
In this paper, we study the full power spectral range of the bright
LMXB and Z source GX~340+0 in order to further investigate the
similarities between the atoll sources and the Z sources, and to help
constrain models concerning the formation of the different QPOs. We
report on the discovery of two new components in the power spectra of
GX~340+0 with frequencies less than 40 Hz when the source is on the
left part of the HB. We also discuss the properties of the NBO, and
those of the kHz QPOs.

\section{Observations and analysis}
\label{analysis}
\noindent
The Z source GX\,340+0 was observed 24 times in 1997 and 1998 with the
proportional counter array (PCA; Jahoda et al. 1996) on board the {\em
RXTE} satellite (Bradt, Rothschild \& Swank 1993). A log of the
observations is presented in Table \ref{obs_log}. Part of these data
(observations 1, 9--18) was used by Jonker et al. (1998) in the
discovery of the kHz QPOs in GX~340+0. The total amount of good data
obtained was $\sim$390 ksec. During $\sim$19\% of the time only 3 or 4
of the 5 PCA detectors were active. 
\begin{deluxetable}{lllll}
\tablecaption{Log of the observations. \label{obs_log}}
\startdata

Number & Observation & Date \& & Total on source \nl
       &    ID       & Start time (UTC) & observing time (ksec.)\nl                         
\tableline
1 & 20054-04-01-00       &   1997-04-17             13:26:21 & 19.8\nl
\tableline
2 & 20059-01-01-00       &   1997-06-06             06:05:07 & 34.7\nl
3 & 20059-01-01-01       &   1997-06-06             21:39:06 & 8.1 \nl
4 & 20059-01-01-02       &   1997-06-07             11:15:05 & 22.1 \nl
5 & 20059-01-01-03       &   1997-06-07             23:48:56 & 21.6 \nl
6 & 20059-01-01-04       &   1997-06-08             07:51:04 & 22.9 \nl
7 & 20059-01-01-05       &   1997-06-09             00:09:03 & 17.5 \nl
8 & 20059-01-01-06       &   1997-06-10             01:22:46 & 22.0\nl
\tableline
9 & 20053-05-01-00       &   1997-09-21             01:04:06 & 17.5\nl
10 & 20053-05-01-01       &   1997-09-23            04:09:30  & 11.5\nl
11 & 20053-05-01-02       &   1997-09-25             01:30:29 & 8.4 \nl
12 & 20053-05-01-03       &   1997-09-25             09:37:51 & 19.3 \nl
13 & 20053-05-02-00       &   1997-11-01             22:38:58 & 9.5\nl
14 & 20053-05-02-01       &   1997-11-02             03:32:07 & 9.0\nl
15 & 20053-05-02-02       &   1997-11-02             19:42:00 & 12.7\nl
16 & 20053-05-02-03       &   1997-11-03             01:50:07 & 13.9\nl
17 & 20053-05-02-04       &   1997-11-04             01:59:34 & 11.1\nl
18 & 20053-05-02-05       &   1997-11-04             16:18:27 & 7.2\nl
\tableline
19 & 30040-04-01-00       &   1998-11-13             23:52:00 & 16.7\nl
20 & 30040-04-01-01       &   1998-11-14             13:55:00 & 17.3 \nl
21 & 30040-04-01-02       &   1998-11-14             21:03:00 & 28.1 \nl
22 & 30040-04-01-03       &   1998-11-15             13:48:00 & 17.1 \nl
23 & 30040-04-01-04       &   1998-11-15             20:57:00 & 17.0\nl
24 & 30040-04-01-05       &   1998-11-15             09:53:00 & 2.6\nl
\enddata
\end{deluxetable}
\par
\noindent
The data were obtained in various modes, of which the Standard 1 and
Standard 2 modes were always active. The Standard 1 mode has a
time resolution of 1/8 s in one energy band (2--60 keV). The
Standard 2 mode has a time resolution of 16 s and the effective 2--60
keV PCA energy range is covered by 129 energy channels. In addition, high
time resolution data (with a resolution of 244 $\mu$s or better for the
2--5.0 keV band and with a resolution of 122 $\mu$s or better for the
5.0--60 keV range) were obtained for all observations.  \par
\noindent
For all observations except observation 1, which had only 4 broad
energy bands, and observation 22, for which technical problems with
the data occurred, we computed power spectra in five broad energy
bands (2--5.0, 5.0--6.4, 6.4--8.6, 8.6--13.0, 13.0--60 keV) with a
Nyquist frequency of 256 Hz dividing the data in segments of 16 s length
each. We also computed power spectra for all observations using 16 s
data segments in one combined broad energy band ranging from 5.0--60
keV with a Nyquist frequency of 4096 Hz.  \par
\noindent
To characterize the properties of the low-frequency part (1/16--256
Hz) of the power spectrum we experimented with several fit functions
(see Section ~\ref{result}) but finally settled on a fit function that
consisted of the sum of a constant to represent the Poisson noise, one
to four Lorentzians describing the QPOs, an exponentially cut-off
power law component, $P\propto\nu^{-\alpha} exp(-\nu/\nu_{cut})$ to
describe the low frequency noise (LFN), and a power law component to
represent the very low frequency noise (VLFN) when the source was on
the NB.  \par
\noindent
To describe the high frequency part (128 to 4096 Hz or 256 to 4096 Hz)
of the power spectrum we used a fit function which consisted of the
sum of a constant and a broad sinusoid to represent the dead-time
modified Poisson noise (Zhang et al. 1995), one or two Lorentzian
peaks to represent the kHz QPOs, and sometimes a power law to fit the
lowest frequency part ($<$ 150 Hz). The PCA setting concerning the
very large event window (Zhang et al. 1995; van der Klis et al. 1997)
was set to 55 $\mu$s. Therefore, its effect on the Poisson noise was
small and it could be incorporated into the broad sinusoid.  The
errors on the fit parameters were determined using $\Delta\chi^2$=1.0
($1 \sigma$ single parameter). The 95\% confidence upper limits were
determined using $\Delta\chi^2$=2.71.  \par
\begin{figure*}[]
\centerline{\psfig{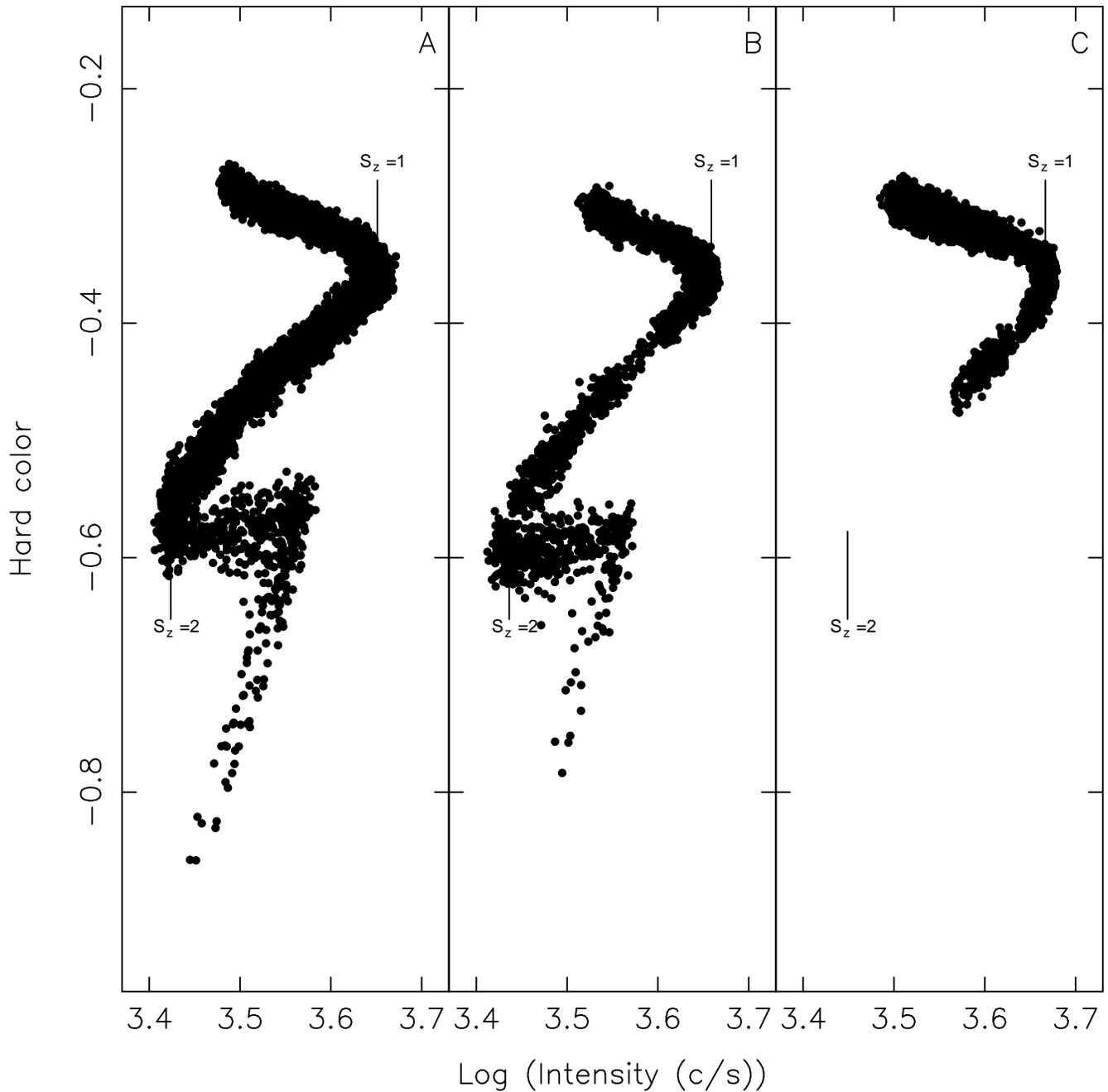}}
\figcaption{Hardness-intensity diagrams for observations 1 and 9--18
(A), 2--8 (B), and 19--24 (C) (see Table \ref{obs_log}). The hard
color is defined as the logarithm of the 9.7--16.0/6.4--9.7 keV count
rate ratio. The intensity is defined as the three-detector count rate
measured in the 2--16.0 keV band. The data were background subtracted
but no dead-time correction was applied. The dead-time correction
factor was less than 1.5\%.
\label{fig_HIDs}}
\end{figure*}

\noindent
We used the Standard 2 data to compute hardnesses and intensities from
the three detectors that were always active. Figure ~\ref{fig_HIDs}
shows three HIDs; one (A) for observations 1 and 9--18 combined (data
set A), one (B) for observation 2--8 combined (data set B), and one
(C) for observation 19--24 combined (data set C). The observations
were subdivided in this way because the hard vertex, defined as the
HB--NB intersection, is at higher intensities in data set C than in
data set A. The hard vertex of data set B falls at an intermediate
intensity level.  

\par
\noindent
We assigned a value to each power spectrum according to the position
of the source along the Z track using the $\rm{S_z}$ parameterization
(Dieters \& van der Klis 1999, Wijnands et al. 1997) applied to each
HID separately. In this parametrization, the hard vertex (defined as
the HB-NB intersection) is assigned the value $\rm{S_z}$ = 1.0 and the
soft vertex (defined as the NB--FB intersection) is assigned
$\rm{S_z}$ = 2.0. Thus, the distance between the hard and soft vertex
defines the length scale along each branch. Since for HID C we only
observed part of the Z track, we used the position of the soft vertex
of HID A in HID C. From the fact that the soft vertex of the HID B was
consistent with that from HID A, we conclude that the error introduced
by this is small.\par
\begin{figure*}[]
\centerline{\psfig{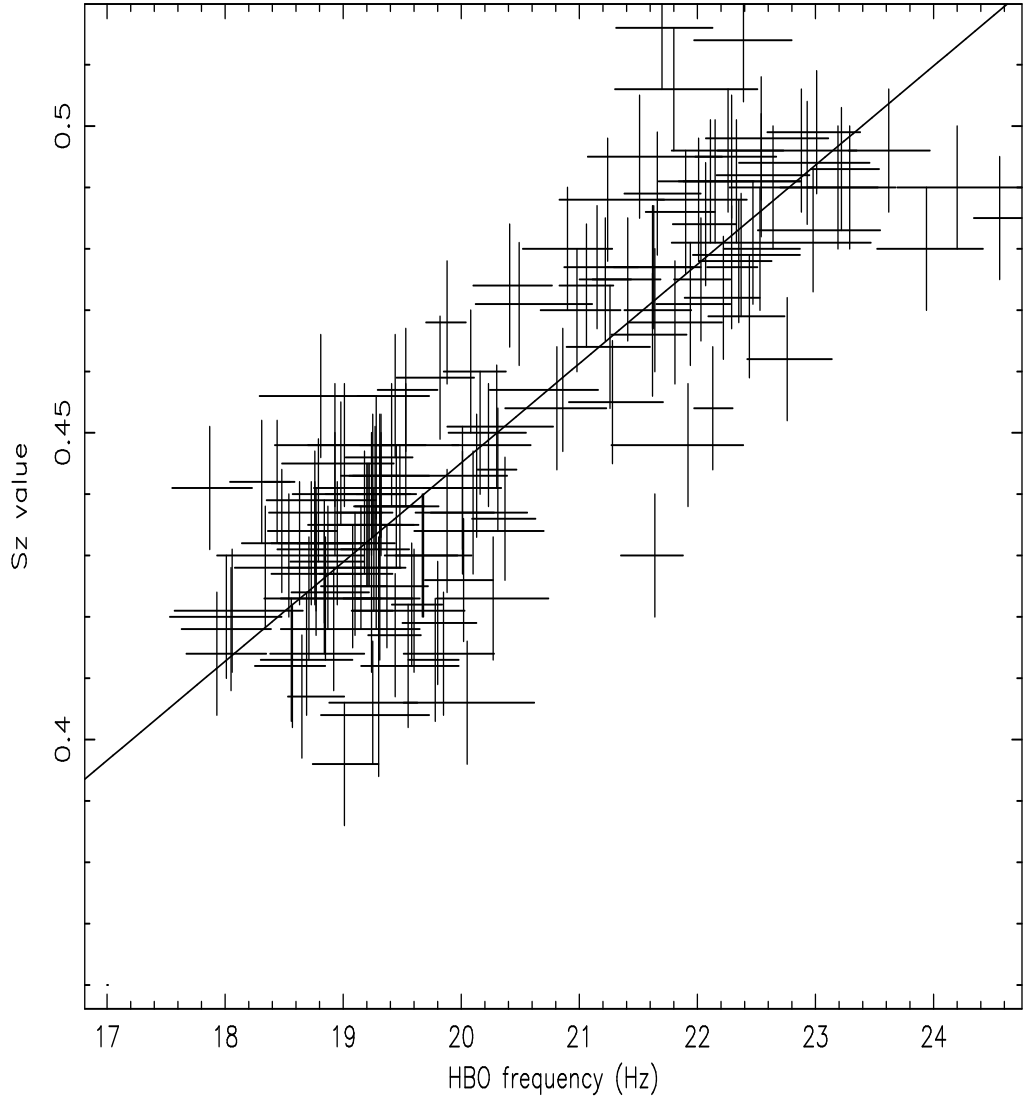}}
\figcaption{The $\rm{S_z}$ value of $\sim $150 individual 16 s length
power spectra from observation 1 plotted against their fitted HBO
frequency. The line represents the best linear fit. The $\chi^2_{red}$
is 1.75 for 144 degrees of freedom, the linear-correlation
coefficient is 0.83.
\label{HBO_vs_Sz}}
\end{figure*}

\noindent
The shifts in the position of the hard vertex prevented us from
selecting the power spectra according to their position in an HID of
all data combined. We selected the power spectra according to the
$\rm{S_z}$ value in each of the three separate Z tracks, since Jonker
et al. (1998) showed that for GX~340+0 the frequency of the HBO is
better correlated to the position of the source relative to the
instantaneous Z track than to its position in terms of coordinates in
the HID. The power spectra corresponding to each $\rm{S_z}$ interval
were averaged. However, employing this method yielded artificially
broadened HBO peaks, and sometimes the HBO profile even displayed
double peaks. The reason for this is that in a typical $\rm{S_z}$
selection interval of 0.05 the dispersion in HBO frequencies well
exceeds the statistical one, as shown in Figure ~\ref{HBO_vs_Sz}.
While the relation between $\rm{S_z}$ and HBO frequency is roughly
linear, the spread is large.
\par
\noindent
For this reason, when the HBO was detectable in the 5.0--60 keV power
spectra, we selected those power spectra according to HBO frequency
rather than on $\rm{S_z}$ value. In practice, this was possible for
all data on the HB. To determine the energy dependence of the
components, the 2--5.0, 5.0--6.4, 6.4--8.6, 8.6--13.0, 13.0--60 keV
power spectra were selected according to the frequency of the HBO peak
in the 2--60 keV power spectrum, when detectable. \par
\begin{figure*}[]
\centerline{\psfig{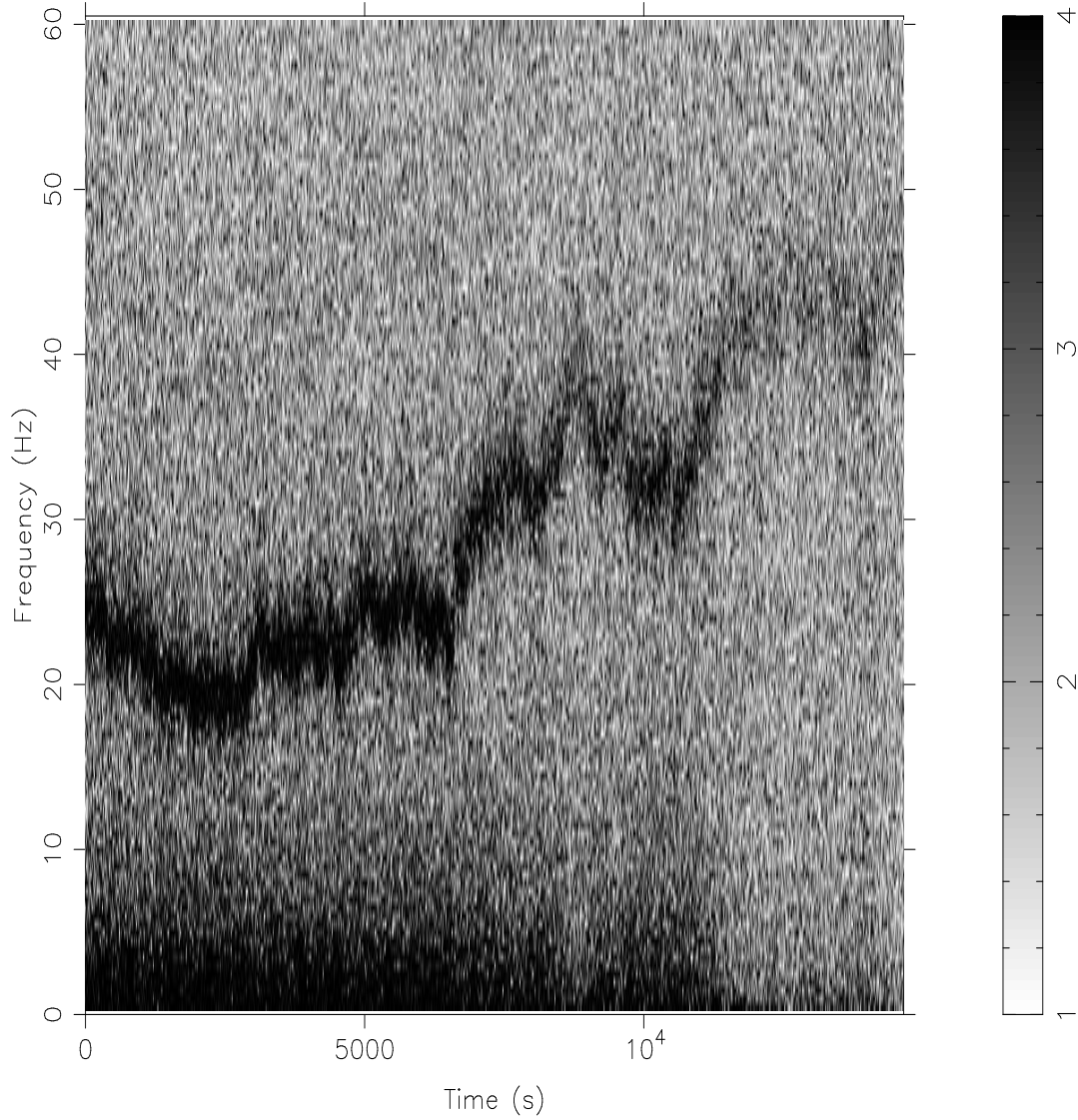}}
\figcaption{The dynamical power spectrum of part of observation 1
showing the 1/16--60 Hz range, with a frequency resolution of 0.5 Hz
in the energy band 5--60 keV. The grey scale represents the Leahy
normalized power (Leahy et al. 1983). Data gaps have been omitted for
clarity. Clearly visible is the HBO at $\sim18$ to $\sim50$ Hz and the
LFN component at low ($<10$ Hz) frequencies.
\label{dynspec}}
\end{figure*}

\noindent
The HBO frequency selection proceeded as follows. For each observation
we constructed a dynamical power spectrum using the 5--60 keV or 2--60
keV data (see above), showing the time evolution of the power spectra
(see Fig.~\ref{dynspec}). Using this method, we were able to trace the
HBO frequency in each observation as a function of time. We determined
the maximum power in 0.5 Hz bins over a range of 2 Hz around the
manually identified QPO frequency for each power spectrum, and adopted
the frequency at which this maximum occurred as the HBO frequency in
that power spectrum. This was done for each observation in which the
HBO could be detected.  
\par
\noindent
The 18--52 Hz frequency range over which the HBO was detected was
divided in 16 selection bins with widths of 2 or 4 Hz, depending on
the signal to noise level. For each selection interval the power
spectra were averaged, and a mean $\rm{S_z}$ value was determined.\par
\noindent
The HBO selection criteria were applied to all data along the HB and
near the hard vertex. When the source was near the hard vertex, on the
NB, or the FB, we selected the power spectra according to the
$\rm{S_z}$ value. An overlap between the two methods occurred for data
near the hard vertex; both selection methods yielded the same results
for the fit parameters to well within the statistical errors (see
Section \ref{result}). Separately, for each set of observations (A,B,
and C) we also determined the kHz QPO properties according to the
$\rm{S_z}$ method.

\section{Results}
\label{result}
\label{fitf}
\noindent
Using the fit function described by Jonker et al. (1998) which
consisted of two Lorentzians to describe the HBO and the second
harmonic of the HBO, and a cut-off power law to describe the LFN noise
component, we obtained poor fits. Compared with Jonker et al. (1998)
we combined more data, resulting in a higher signal to noise
ratio. First we included a peaked noise component (called sub-HBO
component) at frequencies below the HBO, since a similar component was
found by van der Klis et al. (1997) in Sco~X--1. This improved the
$\chi^2_{red}$ of the fit. Remaining problems were that the frequency
of the second harmonic was not equal to twice the HBO frequency
(similar problems fitting the power spectra on the HB of Cyg~X--2 were
reported by Kuulkers, Wijnands, \& van der Klis 1999), and the
frequency of the sub-HBO component varied erratically along the
HB. Inspecting the fit showed that both the fit to the high frequency
tail of the HBO, and the fit to its second harmonic did not represent
the data points very well. Including an additional component in the
fit function representing the high frequency tail of the HBO (called
shoulder component after Belloni et al. [1997] who used this name)
resulted in a better fit to the HBO peak, a centroid frequency of the
HBO second harmonic more nearly equal to twice the HBO frequency, and
a more consistent behavior of the frequency of the sub-HBO component
(which sometimes apparently fitted the shoulder when no shoulder
component was present in the fit function).\par
\noindent
We also experimented with several other fit function components to
describe the average power spectra which were used by other authors to
describe the power spectra of other LMXBs. Using a fit function built
up out of a broken power law, to fit the LFN component, and several
Lorentzians to fit the QPOs after Wijnands et al. (1999a) results in
significantly higher $\chi^2_{red}$ values than when the fit function
described in Section~\ref{analysis} was used ($\chi^2_{red}$= 1.66 for
205 degrees of freedom (d.o.f.) versus a $\chi^2_{red}$= 1.28 with 204
d.o.f.). We also fitted the power spectra using the same fit function
as described in Section~\ref{analysis} but with the frequency of the
sub-HBO component fixed at 0 Hz, in order to test whether or not an
extra LFN-like component centred around 0 Hz was a good representation
of the extra sub-HBO component.  Finally, we tested a fit function
built up out of two cut-off power laws; one describing the LFN and one
either describing the sub-HBO component or the shoulder component, and
three Lorentzians, describing the HBO, its second harmonic, and either
the sub-HBO or shoulder component when not fitted with the cut-off
power law. But in all cases the $\chi^2_{red}$ values obtained using
these fit functions were significantly higher (for the 24--26 Hz
selection range values of 1.52 for 205 d.o.f., 1.62 for 204 d.o.f.,
and 2.00 for 205 d.o.f. were obtained, respectively).  \par
\begin{figure*}[]
\centerline{\psfig{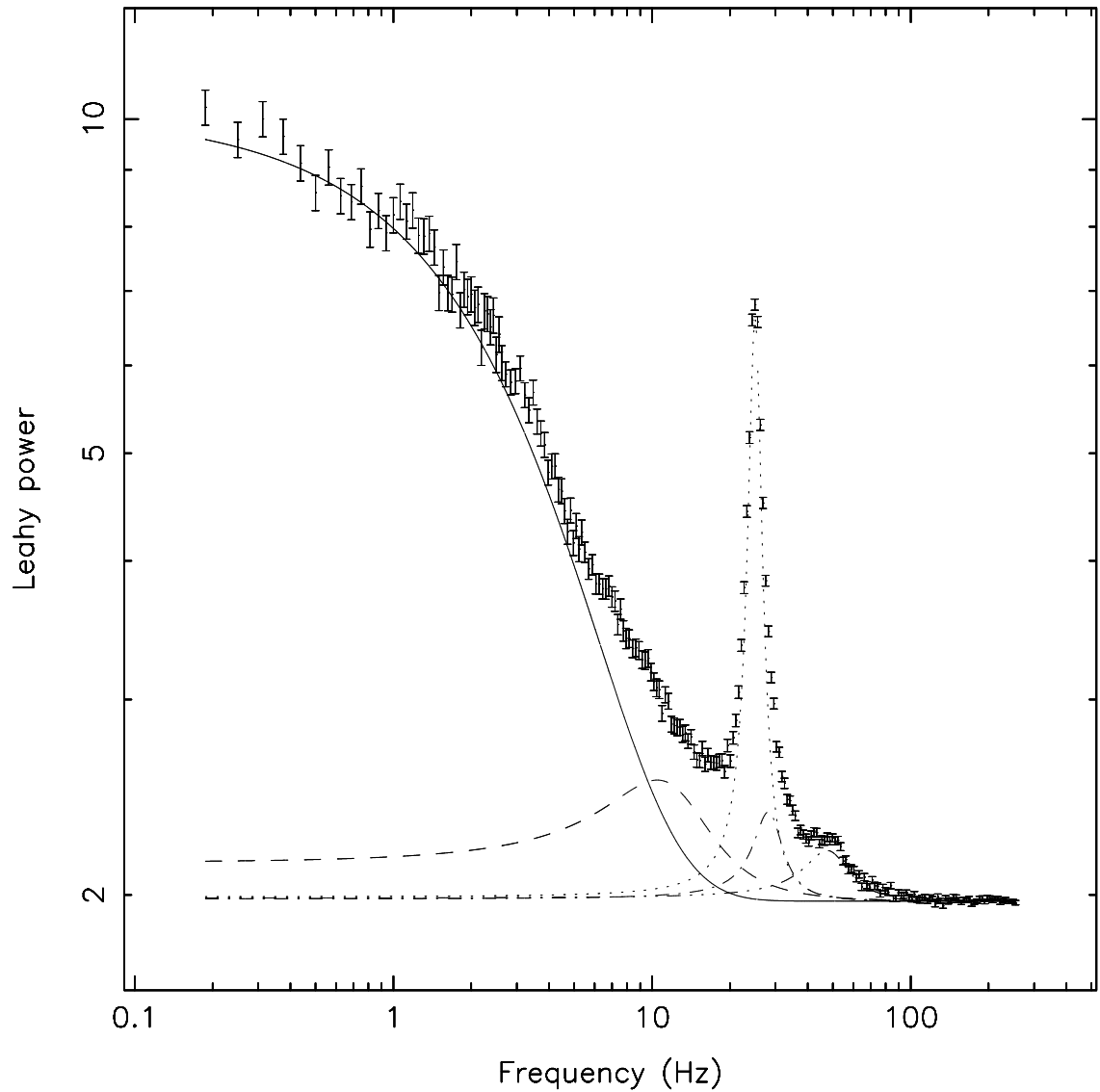}}
\figcaption{Leahy normalized power spectrum showing the different
components used to fit the 5.0--60 keV power spectrum. The full line
represents the LFN component, and the constant arising in the power
spectrum due to the Poisson counting noise; the dashed line represents
the sub-HBO component; the dotted line represents the HBO; the
dashed-dotted line represents the shoulder component; and the
dash-three dots-dash line represents the harmonic of the
HBO.
\label{separate_comps}}
\end{figure*}

\noindent
Settling on the fit function already described in
Section~\ref{analysis}, we applied an F-test (Bevington \& Robinson
1992) to the $\chi^2$ of the fits with and without the extra
Lorentzian components to test their significance. We derived a
significance of more than 8 $\sigma$ for the sub-HBO component, and a
significance of more than 6.5 $\sigma$ for the shoulder component, in
the average selected power spectrum corresponding to HBO frequencies
of 24 to 26 Hz. In Figure~\ref{separate_comps} we show the
contribution of all the components used to obtain the best fit in this
power spectrum.
\par
\begin{figure*}[]
\centerline{\psfig{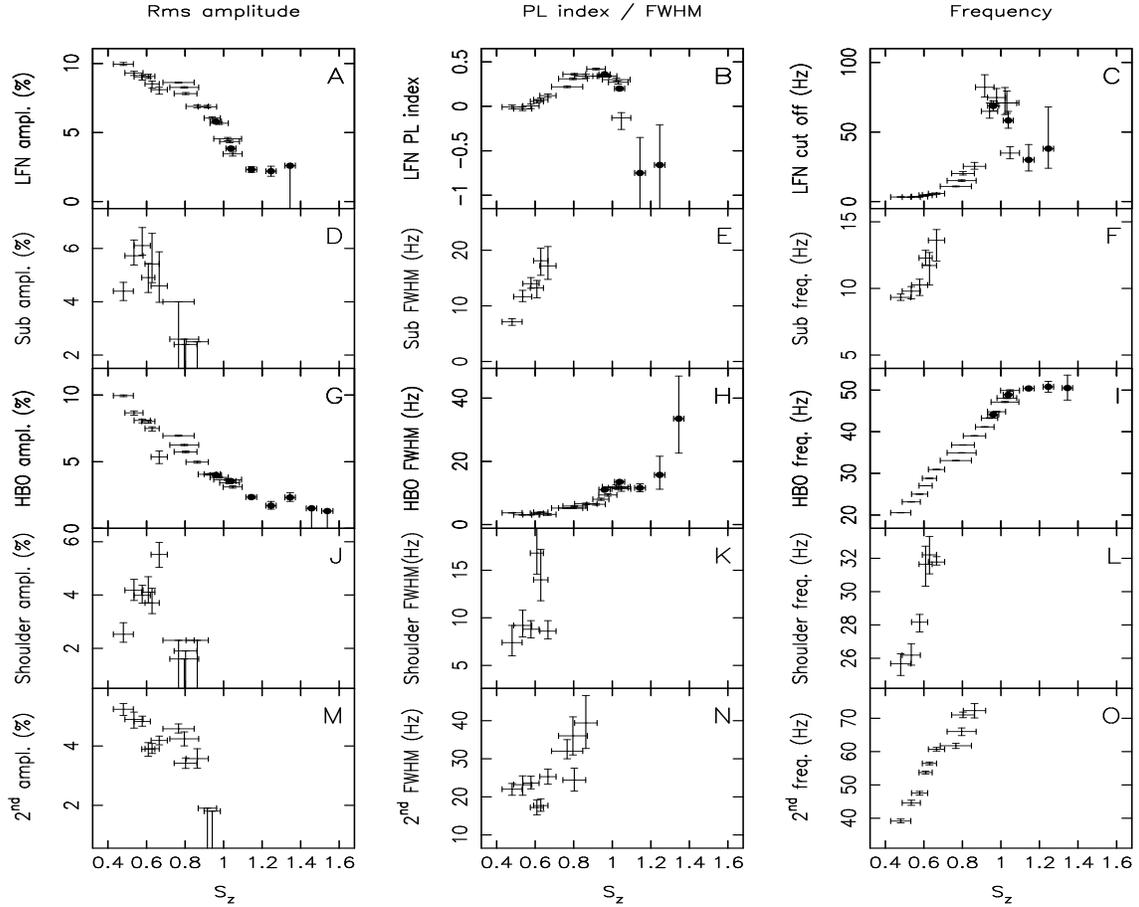}}
\figcaption{(A) Rms amplitude of the low-frequency noise (LFN); (B)
power law index of the LFN; (C) cut-off frequency of the LFN; (D) rms
amplitude of the noise component at frequencies below the HBO
frequency (sub-HBO component); (E) FWHM of the sub-HBO component; (F)
frequency of the sub-HBO component; (G) rms amplitude of the HBO; (H)
FWHM of the HBO; (I) frequency of the HBO; (J) rms amplitude of the
shoulder component used to describe the HBO; (K) FWHM of the shoulder
component; (L) frequency of the shoulder component; (M) rms amplitude
of the harmonic of the HBO; (N) FWHM of the harmonic; (O) frequency of
the harmonic. The points represent data selected according to the HBO
selection method and the bullets represent the data selected according
to the $\rm{S_z}$ selection method (parameters measured in the 5.0--60
keV band, see text). The two methods overlap starting around
$\rm{S_z}\sim$1.0.
\label{all_low_freq_prop}}
\end{figure*}

\noindent
The properties of all the components used in describing the
low-frequency part of the average power spectra along the HB are given
in Fig.~\ref{all_low_freq_prop} as a function of $\rm{S_z}$. When the
HBO frequency was higher than 32 Hz, the sub-HBO and shoulder
component were not significant. We therefore decided to exclude these
two components from the fit function in the HBO frequency selections
of 32 Hz and higher. When this affected the parameters determined for
the remaining components in the fit function, we mention so. Splitting
the total counts into different photon energies reduced the signal to
noise in each energy band and therefore these effects were more
important in the fits performed to determine the energy dependence of
the parameters.

\subsection{The LFN component}
\noindent
The fractional rms amplitude of the LFN decreased as a function of
$\rm{S_z}$ (Fig~\ref{all_low_freq_prop} A), with values ranging from
10\% to 2.2\% (5.0--60 keV). Upper limits on the LFN component were
calculated by fixing the power law index at 0.0. The power law index
of the LFN component increased from $\sim$ 0 at $\rm{S_z} \sim$ 0.5 to
$\sim$ 0.4 around $\rm{S_z}$= 0.9; when the source moved on to the NB
the index of the power law decreased to values slightly below 0.0
(Fig~\ref{all_low_freq_prop} B). The cut-off frequency of the LFN
component increased as a function of $\rm{S_z}$. For $\rm{S_z}> 1.0$
the cut-off frequency could not be determined with high accuracy
(Fig~\ref{all_low_freq_prop} C). \par
\noindent
The LFN fractional rms amplitude depended strongly on photon energy
all across the selected frequency range. The rms amplitude increased
from 5\% at 2--5.0 keV to more than 15\% at 13.0--60 keV
($\rm{S_z}$=0.48). The power law index, $\alpha$, of the LFN
component was higher at lower photon energies (changing from 0.3--0.5
along the HB at 2--5.0 keV) than at higher photon energies (changing
from $-$0.2--0.2 along the HB at 13.0--60 keV). The cut-off frequency
of the LFN component did not change as a function of photon energy.

\subsection{The HBO component}
\noindent
The fractional rms amplitude of the HBO decreased as a function of
$\rm{S_z}$ (Fig~\ref{all_low_freq_prop} G), with values ranging from
10\% to 1.7\% over the detected range (5.0--60 keV). Upper limits on
the HBO component were determined using a fixed FWHM of 15 Hz.  The
frequency of the HBO increased as a function of $\rm{S_z}$ but for
$\rm{S_z}>$1.0 it was consistent with being constant around 50 Hz
(Fig~\ref{all_low_freq_prop} I).  \par
\noindent
The ratio of the rms amplitudes of the LFN and the HBO component, of
interest in beat frequency models (see Shibazaki \& Lamb 1987)
decreased from $\sim$ 1 at an $\rm{S_z}$ value of 0.48 to $\sim 0.6$
at $\rm{S_z}$ values of 0.8--1.0. The ratio increased again to a value
of $\sim$ 0.9 at $\rm{S_z}=$1.05 when the source was on the NB.\par
\begin{figure*}[]
\centerline{\psfig{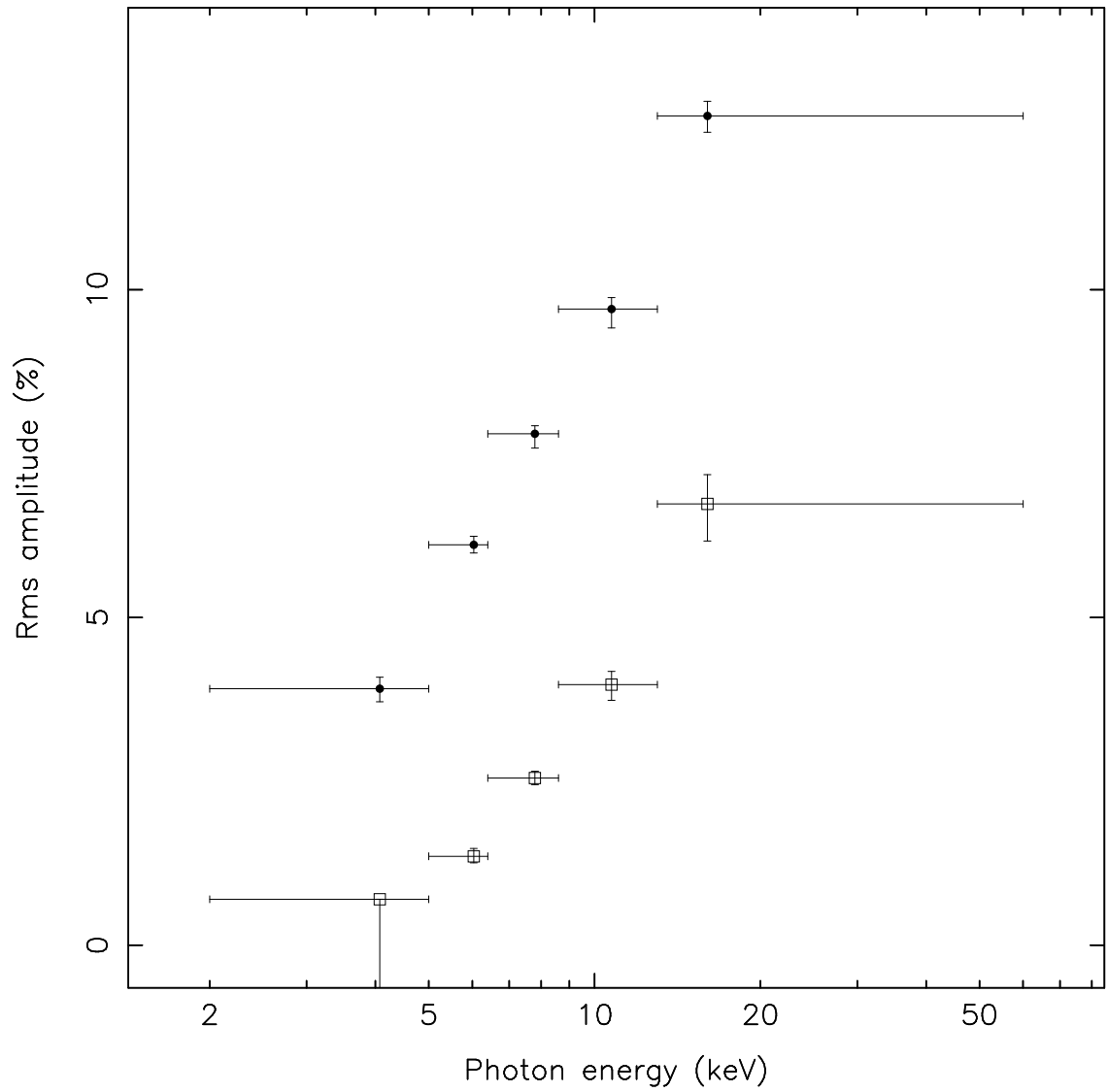}}
\figcaption{The figure shows the typical energy dependence of the rms
amplitude of the HBO (bullets) and NBO (squares) as measured in the
frequency range 26--28 Hz for the HBO, and as measured in the
$\rm{S_z}$ 1.0--1.9 range for the NBO.
\label{en_dep}}
\end{figure*}
\begin{figure*}[]
\centerline{\psfig{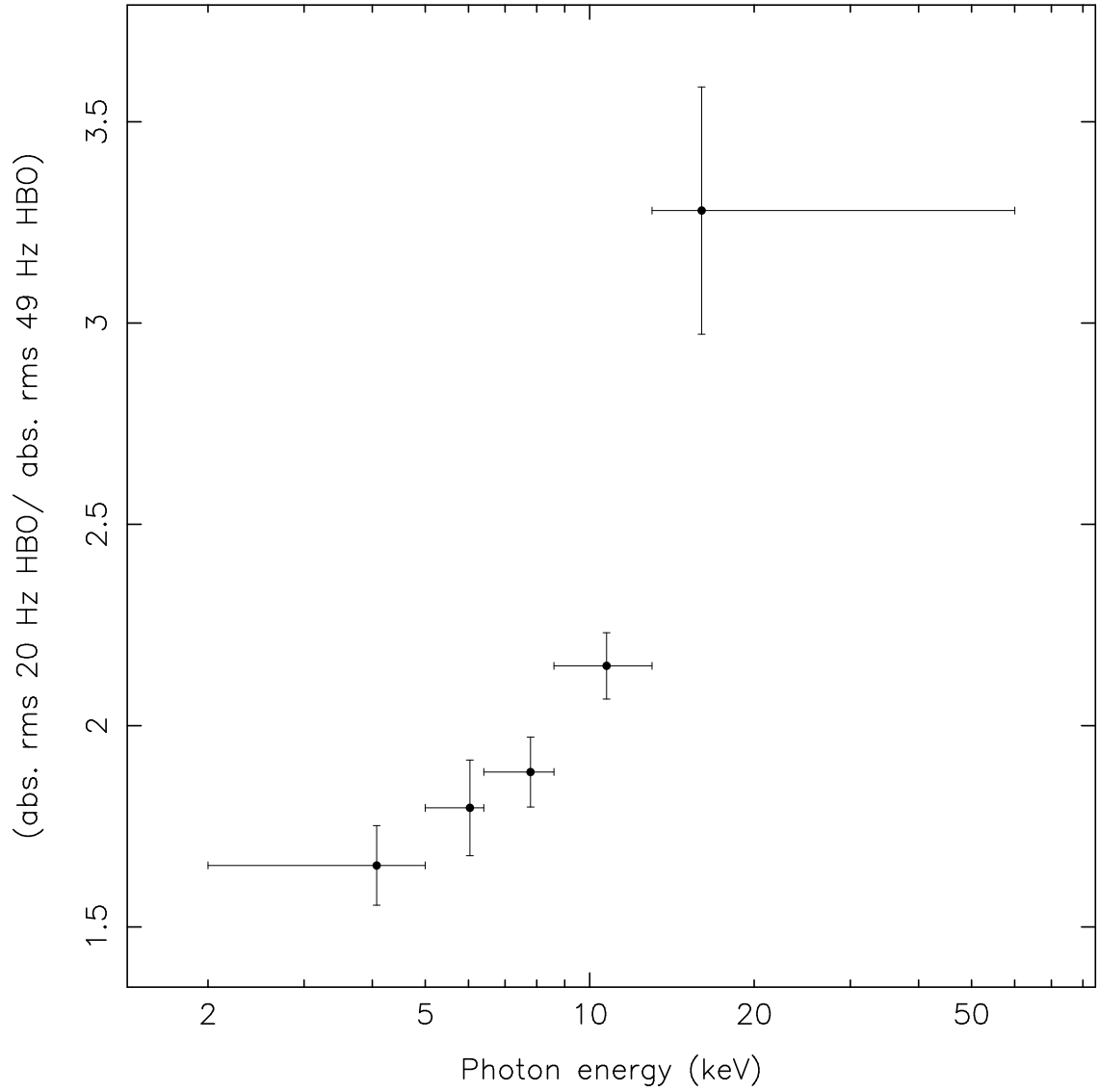}}
\figcaption{The absolute rms amplitude of the HBO at $\sim$20 Hz
divided by the absolute rms amplitude of the HBO at $\sim$50 Hz as a
function of the photon energy.
\label{rmsratio}}
\end{figure*}

\noindent
The HBO rms amplitude depended strongly on photon energy all across
the selected frequency range. The rms amplitude increased from 5\% at
2--5.0 keV to $\sim$ 16\% at 13.0--60 keV (at $\rm{S_z}$=0.48) (see
Figure~\ref{en_dep} [dots] for the HBO energy dependence in the 26--28
Hz range). The increase in fractional rms amplitude of the HBO towards
higher photon energies became less as the frequency of the HBO
increased. At the highest HBO frequencies the HBO is relatively
stronger in the 8.4--13.0 keV band than in the 13.0--60 keV band. The
ratio between the fractional rms amplitude as a function of photon
energy of the HBO at lower frequencies and the fractional rms
amplitude as a function of photon energy of the HBO at higher
frequencies is consistent with a straight line with a positive
slope. The exact fit parameters depend on the HBO frequencies at which
the ratios were taken. This behavior was also present in absolute rms
amplitude ($\equiv \rm{fractional\,rms\,amplitude} * I_{x}$, where
$\rm{I_{x}}$ is the count rate), see Fig.~\ref{rmsratio}. So, this
behavior is caused by actual changes in the QPO spectrum, not by
changes in the time-averaged spectrum by which the QPO spectrum is
divided to calculate the fractional rms spectrum of the HBO. The FWHM
and the frequency of the HBO were the same in each energy band.

\subsection{The second harmonic of the HBO}
\begin{figure*}[]
\centerline{\psfig{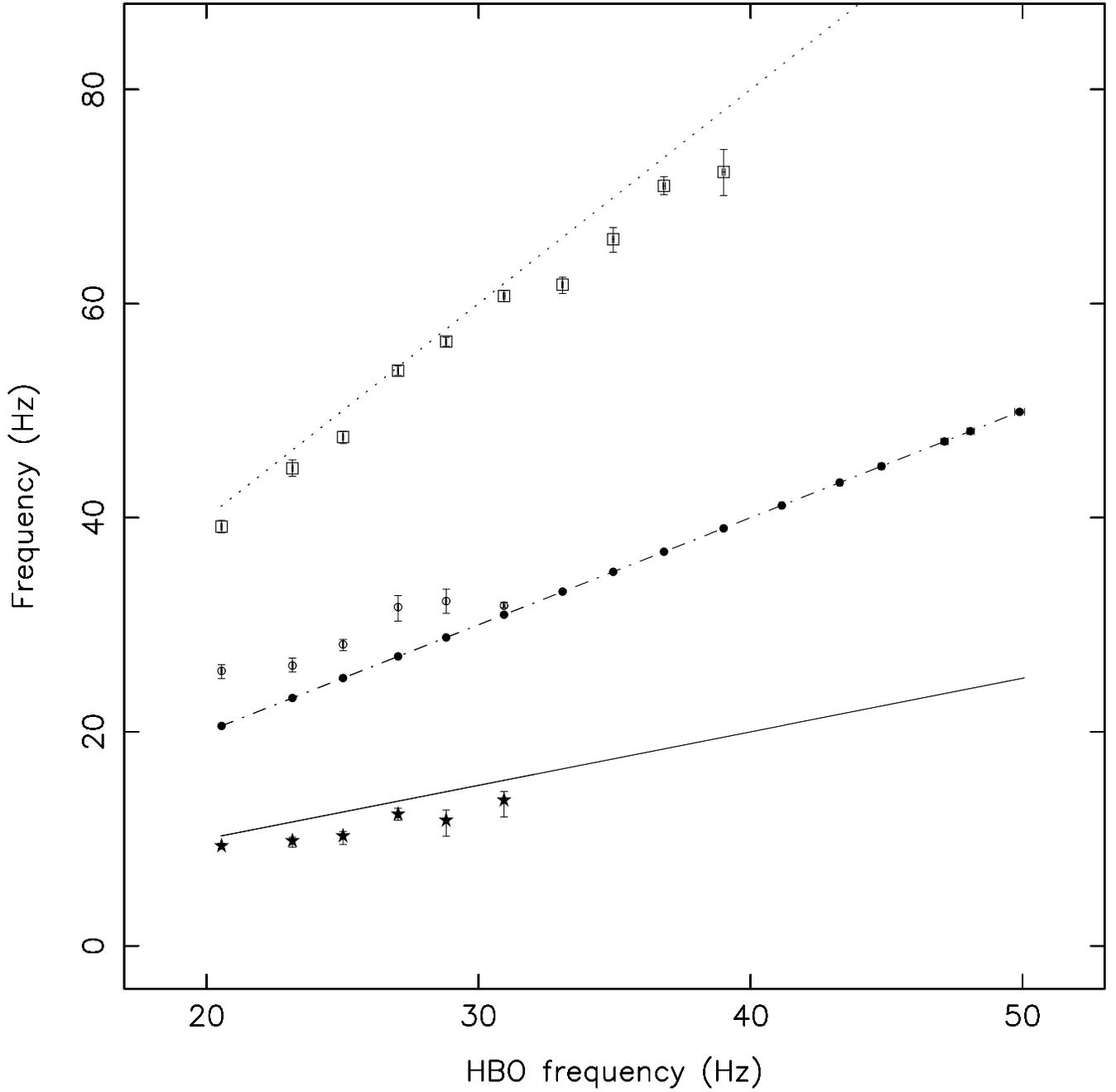}}
\figcaption{The frequencies of the four Lorentzian components used to
describe the average 5.0--60 keV power spectra, as a function of the
HBO frequency. Shown are from low frequencies to high frequencies; the
sub-HBO component (stars), the HBO (bullets), the shoulder component
(open circles), and the second harmonic of the HBO (squares). The
solid line represents the relation $\nu = 0.5 * \nu_{HBO}$, the
dashed-dotted line represents $\nu = 1.0 * \nu_{HBO}$, and the dotted
line represents $\nu = 2.0 * \nu_{HBO}$. Errors in the HBO frequency
are in some cases smaller than the symbols.
\label{freqs}}
\end{figure*}

\noindent
The rms amplitude of the second harmonic of the HBO decreased as a
function of $\rm{S_z}$ (Fig~\ref{all_low_freq_prop} M) from 5.2\% to
3.6\% (5.0--60 keV). Upper limits on the second harmonic of the HBO
were derived using a fixed FWHM of 25 Hz.  The frequency of the second
harmonic of the HBO was consistent with being twice the HBO frequency
when the sub-HBO and the shoulder component were strong enough to be
measured (see Fig~\ref{all_low_freq_prop} O, and
Fig.~\ref{freqs}). When these two extra components could not be
determined significantly, due to the limited signal to noise, and we
therefore omitted them from the fit function (as explained above),
the frequency of the second harmonic of the HBO was clearly less than
twice the HBO frequency (see Fig.~\ref{freqs}).
\par
\noindent
The rms amplitude of the second harmonic of the HBO was also energy
dependent. Its rms amplitude increased from less than 4\% in the
2--5.0 keV, to more than 9\% in the 8.6--13 keV band. The FWHM of the
second harmonic varied erratically in the range of 10--50 Hz. This is
not necessarily a property of the second harmonic since the HBO
shoulder component which was not significant by themselve was omitted
from the fit function. This may have influenced the fit to the FWHM of
the second harmonic when it was weak. Its frequency was consistent
with being the same in each energy band.

\subsection{The sub-HBO component}
\noindent
The centroid frequency (Fig~\ref{all_low_freq_prop} F) and FWHM
(Fig~\ref{all_low_freq_prop} E) of the Lorentzian at sub-HBO
frequencies increased from 9.3$\pm$0.3 Hz to 13.6$\pm$1.0 Hz and from
7.1$\pm$0.6 Hz to 18$\pm$3 Hz, respectively, as the source moved up
the HB from $\rm{S_z}=$ 0.48 to 0.67. The rms amplitude of this
component did not show a clear relation with $\rm{S_z}$; its value was
consistent with being constant around 5\% (Fig~\ref{all_low_freq_prop}
D). Upper limits on the sub-HBO component were determined using a
fixed FWHM of 15 Hz. The frequency of the sub-HBO component is close
to half the frequency of the HBO component. The fact that the ratio
between the HBO frequency and the sub-HBO frequency is not exactly 2
but $\sim2.2$ may be accounted for by the complexity of the data
and therefore its description.  \par
\noindent
We detected the sub-HBO component in the three highest energy bands
that we defined over an $\rm{S_z}$ range from 0.48--0.65. Its rms
amplitude is higher in the highest energy band ($\sim$7\% in 13.0--60
keV, and less than 5\% in 6.4--8.6 keV) and decreased as a function of
$\rm{S_z}$, while the FWHM and the frequency increased from 6--12 Hz,
and 9--15 Hz, respectively.

\subsection{The HBO shoulder component}
\noindent
At an $\rm{S_z}$ value of 0.48 (the left most part of the HB) the
frequency of the shoulder component was higher than the frequency of
the HBO, and the frequency separation between them was largest
(Fig~\ref{all_low_freq_prop} L and Fig.~\ref{freqs}). Both the
frequency of the shoulder and the HBO increased when the source moved
along the HB, but the frequency difference decreased. The FWHM of the
shoulder component increased from 7$\pm$2 Hz to 17$\pm$3 Hz and then
decreased again to 8.6$\pm$1.0 Hz as the frequency of the HBO peak
increased from 25.7$\pm$0.7 Hz to 30.9$\pm$0.4 Hz
(Fig~\ref{all_low_freq_prop} K). From an $\rm{S_z}$ value of 0.61 to
0.67 the frequency was consistent with being constant at a value of 32
Hz. The rms amplitude was consistent with being constant around 4\%
(5.0--60 keV), over the total range where this component could be
detected (Fig~\ref{all_low_freq_prop} J), but the data is also
consistent with an increase of fractional rms amplitude with
increasing HBO frequency. Upper limits on the HBO shoulder component
were determined using a fixed FWHM of 7 Hz.  \par
\noindent
In the various energy bands the HBO shoulder component was detected
seven times in total; once in the 2--5.0 keV band, three times in the
6.4--8.6 keV band, and three times in the in the 8.6--13.0 keV band,
with rms amplitudes increasing from $\sim$3\% in the 2--5.0 keV band
to $\sim$6\% in the 8.6--13.0 keV band, and a FWHM of $\sim$10
Hz. Upper limits of the order of 3\%--4\%, and of 5\%--7\% were
derived in the two lowest and three highest energy bands considered,
respectively. These are comparable with or higher than the rms
amplitudes of this component determined in the 5--60 keV band.

\subsection{The NBO component}
\noindent
The NBOs were not observed when the source was on the HB, with an
upper limit of 0.5\% just before the hard vertex (for an $\rm{S_z}$
value of 0.96). They were detected along the entire NB and they
evolved into a broad noise component on the FB. The properties are
listed in Table~\ref{nbo_tab}. The rms amplitude of the NBO gradually
increased while the source moved from the upper NB to the middle part
of the NB where the rms amplitude is highest. On the lower part of the
NB the NBO rms amplitude gradually decreased. Upper limits on the NBO
components were determined using a fixed FWHM of 5 Hz.
\begin{table*}
\footnotesize
\begin{center}
\begin{tabular}{lllllllll}
\tableline
\tableline
 & NBO & & & shoulder & & & Total &\\
\tableline
$\rm{S_z}$ & $\nu_{NBO}$ & FWHM  & Rms &
$\nu_{shoulder}$& FWHM & Rms & $\nu_{weighted}$ & Total rms\\
        &    (Hz)              & (Hz) &   (\%)       & 
(Hz) & (Hz) &  (\%) & (Hz) & (\%)\\
\tableline
$0.96\pm0.03$ & $8^{\em a}$ & $5^{\em a}$ & $<0.5$ & -- & -- & --& --& -- \\
$1.04\pm0.03$ & $6.2\pm0.4$ & $10\pm2$ & $1.8\pm0.2$& $8^{\em a}$
& $5^{\em a}$ & $<1$ & -- & --\\
$1.14\pm0.03$ & $5.74\pm0.07$ & $6.3\pm0.3$ & $3.4\pm0.1$ & $8^{\em
a}$ & $5^{\em a}$ & $<2$& -- & --\\
$1.25\pm0.03$ & $4.98\pm0.07$ & $2.2\pm0.2$ & $2.8\pm0.3$ &
$6.7\pm0.3$ & $5.0\pm0.4$ & $3.1\pm0.3$& $5.27\pm0.07$ & $4.2\pm0.8$ \\
$1.35\pm0.03$ & $5.40\pm0.05$ & $2.2\pm0.2$ & $4.1\pm0.3$ &
$7.5\pm0.7$ & $5.4\pm0.7$ & $2.6\pm0.4$ &
$5.74\pm0.05$ & $4.9\pm1.0$ \\
$1.46\pm0.03$ & $5.65\pm0.05$ & $2.5\pm0.2$ & $4.0\pm0.2$ &
$8.4\pm1.1$ & $8.5\pm1.7$ & $2.5\pm0.4$ &
$5.87\pm0.05$ & $4.7\pm0.9$ \\
$1.54\pm0.03$ & $5.67\pm0.12$ & $2.9\pm0.5$ &
$3.3\pm0.5$ & $7.5^{+1.4}_{-0.8}$ & $9^{+5}_{-2}$ &
$2.8\pm0.6$ & $5.83\pm0.18$ & $4.3\pm1.6$ \\
$1.65\pm0.03$ & $5.9\pm0.2$ & $5.3\pm0.7$ & $3.2\pm0.1$ & $8^{\em a}$
& $5^{\em a}$& $<2.3$ & -- & --\\
$1.75\pm0.03$ & $6.1\pm0.3$ & $5.2\pm0.9$ &
$2.4\pm0.2$ & -- & -- & -- & -- & --\\
$1.85\pm0.04$ & $7.1\pm0.8$& $6.7\pm1.6$ & $2.1\pm0.2$& -- & -- & --& -- & --\\
$1.95\pm0.04$ & $6.3\pm0.9$ & $5\pm3$ & $1.7\pm0.3$ & --& -- & --& -- & --\\
$2.05\pm0.03$ & $5.6\pm1.4$ & $12\pm4$&$2.6\pm0.4$& -- & -- & --& -- & --\\
$2.15\pm0.03$ & $ 4^{+3}_{-10}$ & $22^{+15}_{-8}$ &
$3.2^{+1.6}_{-0.5}$& -- & -- & --& -- & --\\ 
\tableline
\end{tabular}
\end{center}
\tablenotetext{a}{Parameter fixed at this value}
\caption{Properties of the NBO fitted using one or two Lorentzians, as
a function of $\rm{S_z}$ in the 5.0--60 keV band.\label{nbo_tab}}
\end{table*}
\par
\begin{figure*}[]
\centerline{\psfig{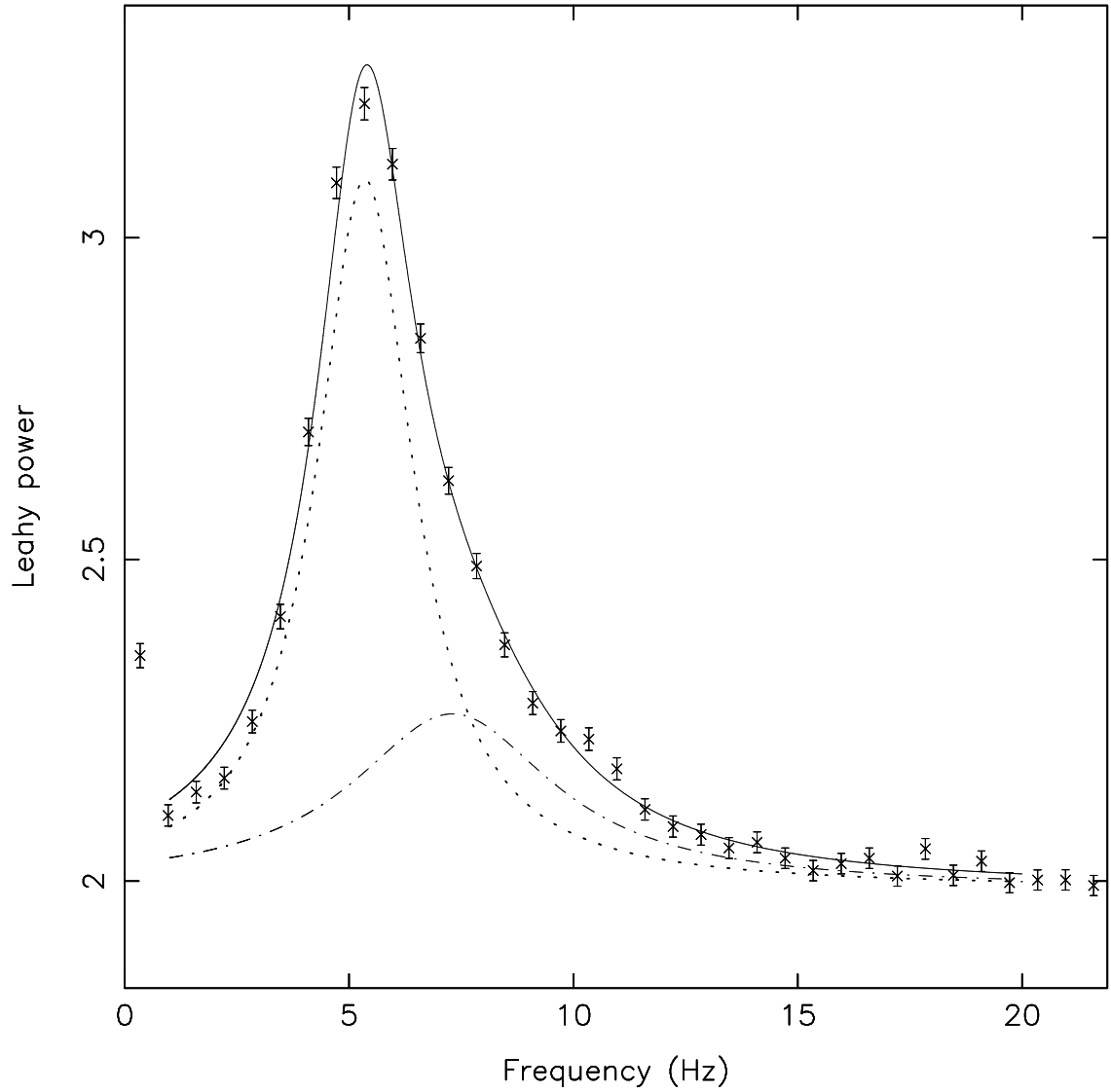}}
\figcaption{Typical Leahy normalized power spectrum on the NB showing
the NBO in the energy band 5.0--60 keV (in the $\rm{S_z}$ 1.2--1.5
range). The asymmetry of the profile is clearly visible; the drawn
line represents the best fit model, using two Lorentzian peaks. The
dotted line and the dash-dotted line represent the two
Lorentzians.
\label{NBO_asym}}
\end{figure*}

\noindent
As the NBO got stronger towards the middle of the NB the profile of
the NBO became detectably asymmetric (see Fig.~\ref{NBO_asym});
between $\rm{S_z}$=1.25 and 1.54 the NBO was fitted using two
Lorentzians. The FWHM of the NBO as a function of the position along
the NB was first decreasing from $~\sim$10 Hz at $\rm{S_z}$=1.038 to
around 2.5 Hz on the middle part of the NB ($\rm{S_z}$ values from
1.25--1.54), and then increased again to $~\sim$ 5 Hz on the lowest
part of the NB.
\par
\noindent
Due to the fact that the NBO profiles had to be fitted using two
Lorentzians in part of the data, the behavior of the NBO frequency as a
function of $\rm{S_z}$ is also not determined
unambiguously. Therefore, we weighted the frequencies of these two
Lorentzians according to one over the square of the FWHM. The FWHM
weighted average of the two centroid frequencies of the two
Lorentzians used to describe the NBO was consistent with a small
increase as a function of $\rm{S_z}$ from $5.27\pm0.07$ Hz at
$\rm{S_z}$=1.25 to $5.83\pm0.18$ Hz at $\rm{S_z}$=1.54.\par
\noindent
We combined all power spectra with an $\rm{S_z}$ between 1.0 and 1.9
in order to investigate the energy dependence of the NBO.  The rms
amplitude of the NBO increased as of function of photon energy (see
Figure~\ref{en_dep} [squares]). 

\subsection{KHz QPOs} 
\label{khz_res}
\begin{figure*}[]
\centerline{\psfig{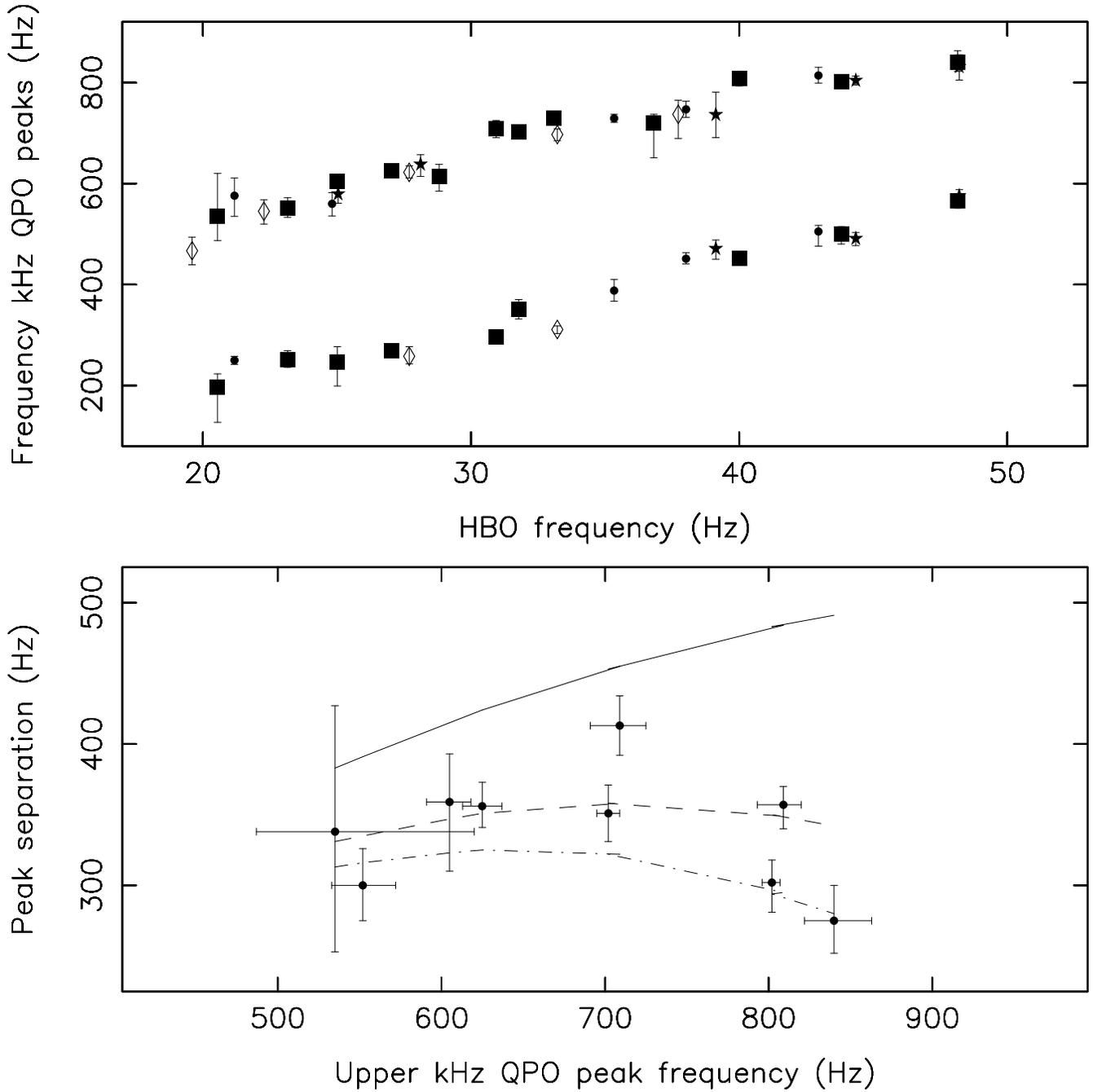}}
\figcaption{Upper panel: Relation between the lower and upper kHz QPO
peak frequencies and the HBO frequency, as measured using all the data
selected according to their HBO frequency in the 5--60 keV energy band
(filled large squares), and using data from observations 1, and 9--18
combined (bullets; see Jonker et al. 1998), data from observations
2--8 combined (stars), and data from observations 19--25 combined
(diamonds) selected according to the $\rm{S_z}$ selection method. The
error bars on the HBO frequency are small compared to the size of the
data points, and are therefore omitted. Lower panel: The peak
separation vs. upper kHz QPO frequency as measured when selected
according to HBO frequency. The solid, dashed, and dash-dotted lines
represent the predicted relations between the peak separation and the
Keplerian frequency in the Stella \& Vietri (1999) model for a neutron
star mass of 1.4, 2.0, and 2.2 $M_{\odot}$,
respectively.
\label{kHz_vs_HBO_two_selection}}
\end{figure*}

\noindent
Using the HBO frequency selection method in all data combined, the
frequency of the kHz QPO peaks increased from $197^{+26}_{-70}$ Hz to
$565^{+9}_{-14}$ Hz and from $535^{+85}_{-48}$ Hz to $840\pm21$
Hz for the lower and upper peak, respectively, while the frequency of
the HBO increased from $20.55\pm0.02$ Hz to $48.15\pm0.08$ Hz. Using
the $\rm{S_z}$ selection method on the three data sets we defined in
Section~\ref{analysis} (Figure~\ref{fig_HIDs}), we found that the
relation between the kHz QPO and the HBO is consistent with being the
same in all three data sets (Fig.~\ref{kHz_vs_HBO_two_selection} upper
panel). The same relation was found when we combined all data and
selected the power spectra according to the HBO frequency.  
\par
\noindent
Upper limits on the kHz QPOs were determined with the FWHM fixed at
150 Hz. When only one of the two kHz QPO peaks was detected the upper
limit on the other peak was determined by fixing the frequency at the
frequency of the detected peak plus or minus the mean difference
frequency between the two peaks, depending on whether the lower or the
upper peak was detected. The properties of the kHz QPOs as determined
in all data combined when selected according to the HBO frequency are
listed in Table ~\ref{khztable}.
\begin{table*}
\footnotesize
\begin{center}
\begin{tabular}{lllllll}
\tableline
\tableline
$\nu_{HBO}$ (Hz) & Upper & FWHM upper & Rms upper & Lower & FWHM lower &
Rms lower \\
 & kHz $\nu$ (Hz) & peak (Hz) & peak (\%) & kHz $\nu$ (Hz) & peak (Hz)
& peak (\%) \\
\tableline
$20.55\pm0.02$ & $535^{+85}_{-48}$ & $334^{+103}_{-172}$ &
$4.2^{+0.8}_{-1.2}$ & $197^{+26}_{-70}$ & $171^{+377}_{-94}$ &
$3.1^{+3.7}_{-0.8}$ \\
$23.16\pm0.02$ & $552\pm20$ & $176\pm43$& $3.3\pm0.4$ &
$252\pm17$ & $125\pm58$ & $2.7\pm0.5$ \\
$25.02\pm0.02$ & $605\pm14$ & $170\pm34$ &
$3.6\pm0.4$ & $246\pm40$ & $208^{+151}_{-85}$ &
$2.8^{+1.0}_{-0.5}$ \\
$27.04\pm0.02$ & $ 625\pm12$ & $156\pm36$ & $3.3\pm0.3$ &
$269\pm11$ & $63\pm25$ & $2.0\pm0.3$\\
$28.81\pm0.03$ & $614\pm27$ & $229^{+118}_{-80}$ &
$3.1\pm0.6$ & $275^{\em a}$ & $150^{\em a}$ & $<1.4$ \\
$30.94\pm0.04$ & $709\pm17$ & $189\pm50$ &
$3.4\pm0.4$ & $296\pm12$ & $100\pm38$ & $2.5\pm0.4$\\
$31.8\pm0.3$ & $702\pm7$ & $83\pm18$ & $2.7\pm0.2$ &
$351\pm19$ & $152^{+92}_{-52}$ & $2.4\pm0.5$ \\
$33.09\pm0.02$ & $729\pm13$ & $119\pm31$ & $2.7\pm0.3$ &
$390^{\em a}$ & $150^{\em a}$ & $<2.4$ \\
$36.82\pm0.05$ & $720^{+17}_{-69}$ & $209^{+389}_{-71}$ &
$3.5^{+2.7}_{-0.5}$ & $382^{\em a}$ & $150^{\em a}$ & $<2.4$ \\
$40.00\pm0.06$ & $809\pm14$ & $86\pm49$ &
$1.9^{+0.2}_{-0.4}$ & $452\pm7$ & $35\pm27$ & $1.2\pm0.3$\\
$43.81\pm0.07$ & $802\pm6$ & $62^{+30}_{-18}$ &
$1.8\pm0.3$ & $500\pm18$ & $73\pm31$ & $1.3\pm0.2$\\
$48.15\pm0.08$ & $840\pm21$ & $109\pm61$ &
$1.2\pm0.3$ & $565\pm12$ & $69^{+46}_{-29}$ & $1.1\pm0.2$\\
\tableline
\end{tabular}
\end{center}
\tablenotetext{a}{Parameter fixed at this value}
\caption{Properties of the kHz QPOs as determined in all 5.0--60 keV
data combined, selected according to the HBO
frequency.\label{khztable}}
\end{table*}
\par
\noindent
The kHz QPO peak separation was consistent with being constant at
339$\pm$8 Hz over the observed kHz QPO range
(Fig.~\ref{kHz_vs_HBO_two_selection} lower panel), but a decrease
towards higher upper peak frequencies similar to that found in
Sco~X--1 (van der Klis et al. 1997), 4U~1608--52 (M\'endez et
al. 1998), 4U~1735--44 (Ford et al. 1998), 4U~1702--429 (Markwardt,
Strohmayer, \& Swank 1999), and 4U~1728--34 (M\'endez \& van der Klis
1999) cannot be excluded. The FWHM of neither the lower nor the higher
frequency kHz QPO peak showed a clear relation with frequency. The rms
amplitude of the lower and upper kHz QPO peak decreased from 3.1\% to
1.1\%, and from 4.2\% to 1.2\%, respectively when the HBO frequency
increased from 20.55 to 48.15 Hz.

\section{Discussion}
\noindent
In the present work we combined all {\em RXTE} data presently
available for the Z source GX~340+0 using our new selection method
based on the frequency of the HBO peak. This allowed us to distinguish
two new components in the low-frequency part of the power
spectrum. \par
\noindent
These two extra components were strongest when the source was at the
lowest count rates on the HB (see Fig.~\ref{fig_HIDs}), between
$\rm{S_z}=$ 0.48--0.73, i.e., at the lowest inferred $\dot{M}$.  The
frequency of one of these components, the sub-HBO component, is close
to half the frequency of the HBO component. The frequency ratio was
consistent with being constant when the frequency of the sub-HBO
changed from 9 to 14 Hz. A similar feature at sub-HBO frequencies has
been reported by van der Klis et al. (1997) in Sco\,X--1. Since the
frequency of this component is close to twice the predicted
Lense-Thirring (LT) precession frequency for rapidly rotating neutron
stars (Stella \& Vietri 1998), we shall discuss the properties of this
component within this framework.  \par
\noindent
The other component we discovered, the HBO shoulder component, was
used to describe the strong excess in power in the HBO profile towards
higher frequencies.  If this shoulder component is related to the HBO
and not to a completely different mechanism which by chance results in
frequencies close to the frequency of the HBO, it can be used to
constrain the formation models of the HBO peak. We demonstrated that
both the HBO and the NBO have a similar asymmetric profile. In the NBO
this was previously noted by Priedhorsky et al. (1986) in Sco~X--1. We
shall consider the hypothesis that the formation of this shoulder is a
common feature of the two different QPO phenomena, even if the two
peaks themselves perhaps occur due to completely different physical
reasons. \par
\noindent
Our results on the kHz QPOs based on more extensive data sets at three
different epochs and using the new HBO selection method are consistent
with those of Jonker et al. (1998). We discuss the properties of the
kHz QPOs within the framework of precessing Keplerian flows (Stella \&
Vietri 1999), the sonic point model (Miller, Lamb, \& Psaltis
1998), and the transition layer model described by Osherovich \&
Titarchuk (1999), and Titarchuk, Osherovich, \& Kuznetsov (1999).

\subsection{Comparison with other observations}
\noindent
In various LMXBs, QPOs have been found whose profiles are clearly not
symmetric. Belloni et al. (1997) showed that for the black hole
candidate (BHC) GS~1124--68 the QPO profiles are asymmetric, with a
high frequency shoulder. Dieters et al. (1998) reported that the 2.67
Hz QPO of the BHC 4U~1630--47 was also asymmetric with a high
frequency shoulder. In the Z source Sco~X--1 the NBO profile was also
found to be asymmetric (Priedhorsky et al. 1986). It is clear that
asymmetric shapes of the QPO profiles are frequently observed in LMXBs
and are not restricted to either the black hole candidates or the
neutron star systems.\par
\noindent
In the BHCs GS~1124--68 (Belloni et al. 1997) and XTE~J1550--564
(Homan et al. 1999) several QPOs were discovered which seem to be
harmonically related in the same way as we report for GX~340+0,
i.e. the third harmonic is not detected, while the first, the second
and the fourth harmonic are. If this implies that these QPOs are the
same, models involving the magnetic field of the neutron star for
their origin could be ruled out. The time lag properties of the
harmonic components of the QPOs in XTE~J1550--564 are complex and
quite distinctive (Wijnands, Homan, \& van der Klis 1999). In GX~340+0
no time lags of the harmonic components could be measured, but the
time lags measured in the HBO in the similar Z source GX~5--1 (Vaughan
et al. 1994) are quite different. \par
\noindent
In order to study in more detail the relationship found by Wijnands \&
van der Klis (1999) between the QPOs and the noise break frequency in
the power spectrum of LMXBs, we fitted the LFN component using a
broken power law. To determine the value for the break frequency we
fixed the parameters of all other components to the values found when
using a cut-off power law to describe the LFN. Wijnands \& van der
Klis (1999a) reported that the Z sources did not fall on the relation
between the break and QPO frequency established by atoll sources and
black hole candidates. They suggested that the Z source LFN is not
similar to the atoll HFN but the noise component found in Sco~X--1 at
sub-HBO frequencies is. By using the centroid frequency of that peaked
noise component as the break frequency instead of the LFN break
frequency, the HBO frequencies did fall on the reported relation. On
the other hand, we find that using the sub-HBO frequency instead of
the HBO frequency together with the LFN break frequency, the Z source
GX~340+0 also falls exactly on the relation.  Therefore, the
suggestion made by Wijnands \& van der Klis (1999a) that the strong
band-limited noise in atoll and Z sources have a different origin is
only one of the two possible solutions to the observed
discrepancy. Our proposed alternative solution is that the Z and atoll
noise components are the same, but that it is the sub-HBO in Z sources
which corresponds to the QPO in atoll sources. An argument in favour
of the noise components in Z and atoll sources being the same is that
the cut-off frequency of the LFN component increased as a function of
$\rm{S_z}$, in a similar fashion as the frequency associated with the
atoll high frequency noise (van der Klis 1995, Ford \& van der Klis
1998, van Straaten et al. 1999). \par
\noindent
Following Psaltis, Belloni, \& van der Klis (1999) we plotted the
sub-HBO frequency against the frequency of the lower-frequency kHz
QPO. The sub-HBO does not fall on the relation found by Psaltis,
Belloni, \& van der Klis (1999) between the frequency of the HBO and
the lower-frequency kHz QPO frequency. Instead the data points fall
between the two branches defined by the HBO-like QPO frequencies
vs. the lower kHz QPO frequency at high frequencies (see Psaltis,
Belloni, \& van der Klis 1999).

\subsection{HBO -- kHz QPO relations}
\subsubsection{Lense-Thirring precession frequency}
\noindent
Stella \& Vietri (1998) recently considered the possibility that the
HBO is formed due to the LT precession of the nodal points of sligthly
tilted orbits in the inner accretion disk, but as they already
mentioned the Z sources GX~5--1 and GX~17+2 did not seem to fit in
this scheme. For reasonable values of I/M, the neutron star moment of
inertia divided by its mass, the observed frequencies were larger by a
factor of $\sim$2 than the predicted ones. Jonker et al. (1998) showed
that for GX~340+0 the predicted frequency is too small by a factor of
3, if one assumes that the higher frequency peak of the kHz QPOs
reflects the Keplerian frequency of matter in orbit around the neutron
star, and that the mean peak separation reflects the neutron star spin
frequency. Using the same assumptions Psaltis et al. (1999) also
concluded that a simple LT precession frequency model is unable to
explain the formation of HBOs in Z sources.  \par
\noindent
Detailed calculations of Morsink \& Stella (1999) even worsen the
situation, since their calculations lower the predicted LT
frequencies. They find that the LT precession frequencies are
approximately a factor of two too low to explain the noise components
at frequencies $\sim$20--35 Hz observed in atoll sources (4U~1735--44,
Wijnands \& van der Klis 1998c; 4U~1728--34, Strohmayer et al 1996,
Ford \& van der Klis 1998). Stella \& Vietri (1998) already put
forward the suggestion that a modulation can be produced at twice the
LT precession frequency if the modulation is produced by the two
points where the inclined orbit intersects the disk plane (although
they initially used this for explaining the discrepancy of a factor of
two between the predicted and the observed LT precession frequencies
for the Z sources).  \par
\noindent
The sub-HBO peaked noise component we discovered could be harmonically
related to the HBO component. If the sub-HBO is the second harmonic of
the fundamental LT precession frequency, as needed to explain the
frequencies in the framework of the LT precession model where the
neutron star spin frequency is approximately equal to the frequency of
the kHz QPO peak separation, the HBO must be the fourth and the
harmonic of the HBO must be the eighth harmonic component, whereas the
sixth and uneven harmonics must be much weaker. This poses strong
(geometrical) constraints on the LT precession process. On the other
hand, if the HBO frequency is twice the LT precession frequency, which
implies a neutron star spin frequency of $\sim$900 Hz (see Morsink \&
Stella 1999), the frequency of the sub-HBO component is the LT
precession frequency, and the frequency of the second harmonic of the
HBO is four times the LT precession frequency. In that case only even
harmonics and the LT precession frequency are observed.

\subsubsection{Magnetospheric beat frequency and radial-flow models}
\label{random}
\noindent
In this section, we discuss our findings concerning the QPOs and the
LFN component in terms of the magnetic beat frequency model where the
QPOs are described by harmonic series (e.g. Shibazaki \& Lamb
1987). \par
\noindent
If the sub-HBO frequency is proven not to be harmonically related to
the HBO, the sub-HBO peak might be explained as an effect of
fluctuations entering the magnetospheric boundary layer
periodically. Such an effect will be strongest at low HBO frequencies
since its power density will be proportional to the power density of
the LFN (Shibazaki \& Lamb 1987). If it is the fundamental frequency
and the HBO its first overtone then the magnetospheric beat frequency
model proposed to explain the HBO formation (Alpar \& Shaham 1985;
Lamb et al. 1985) is not strongly constrained.  \par
\noindent
Within the beat frequency model the high frequency shoulder of the HBO
peak can be explained as a sign of radial drift of the blobs as they
spiral in after crossing the magnetospheric boundary layer (Shibazaki
\& Lamb 1987). Shibazaki \& Lamb (1987) describe another mechanism
which may produce a high frequency shoulder. Interference between the
LFN and the QPO caused by a non uniform phase distribution of the
blobs will also cause the QPO to become asymmetric. This effect will
be strongest when the LFN and the QPO components overlap, as
observed. Finally, an asymmetric initial distribution of frequencies
of the blobs when entering the magnetospheric boundary layer may also
form an asymmetric HBO peak.  \par
\noindent
The changes in the power law index of the LFN as a function of photon
energy can be explained by varying the width or the steepness of the
lifetime distribution of the blobs entering the magnetic boundary
layer (Shibazaki \& Lamb 1987). The decrease in increase of both the
fractional and absolute rms amplitude of the HBO as a function of
energy towards higher frequencies (Fig.~\ref{rmsratio}) also
constrains the detailed physical interactions occurring in the
boundary layer.  \par
\noindent
Fortner et al. (1989) proposed that the NBO is caused by oscillations
in the electron scattering optical depth at the critical Eddington
mass accretion rate. How a high frequency shoulder can be produced
within this model is not clear. Both the HBO and the NBO shoulder
components were detected when the rms amplitude of the HBO and the NBO
was highest. In case of the NBO, this may be a result of the higher
signal to noise. Since the rms amplitude of the NBO shoulder component
is consistent with being $\sim$2/3 of the NBO rms amplitude (see
Table~\ref{nbo_tab}), combining more observations should increase the
range over which this shoulder component is detected, if this ratio is
constant along $\rm{S_z}$. In case of the HBO the two components seem
to merge. While the fractional rms amplitude of the HBO shoulder
component increased that of the HBO decreased. When the fractional rms
amplitudes were comparable, the HBO was fitted with one
Lorentzian. The rms amplitude of both shoulder components increased in
a similar way as the rms amplitudes of the NBO and the HBO with photon
energy. So, the formation of these shoulder components seems a common
feature of both QPO forming mechanisms.

\subsubsection{Radial oscillations in a viscous layer}
In Sco~X--1, Titarchuk, Osherovich, \& Kuznetsov (1999) interpreted
the extra noise component in the power spectra (van der Klis et
al. 1997) as due to radial oscillations in a viscous boundary layer
(Titarchuk \& Osherovich 1999). If the noise component in Sco~X--1 is
the sub-HBO component in GX~340+0, the model of Titarchuk \&
Osherovich (1999) can be applied to the frequencies and dependencies
we found for the sub-HBO component in GX~340+0. Fitting our data to
the relation between the frequency of the extra noise component and
the Keplerian frequency, using the parameters and parametrization
given by Titarchuk, Osherovich, \& Kuznetsov (1999), we obtained a
value of $C_N=15$ for GX~340+0. This value is much larger than the
value obtained for Sco~X--1 (9.76). According to Titarchuk \&
Osherovich (1999) a higher $C_N$ value implies a higher viscosity for
the same Reynold's number.

\subsection{KHz QPOs and their peak separation}
\noindent
Recently, Stella \& Vietri (1999) have put forward a model in which
the formation of the lower kHz QPO is due to the relativistic
periastron precession (apsidal motion) of matter in (near) Keplerian
orbits. The frequency of the upper kHz QPO peaks is the Keplerian
frequency of this material. The peak separation is then equal to the
radial frequency of matter in a nearly circular Keplerian orbit, and
is predicted to decrease as the Keplerian frequency increases and
approaches the predicted frequency at the marginally stable circular
orbit. This model can explain the decrease in peak separation as
observed in various sources (see Section~\ref{khz_res}).  \par
\noindent
Beat frequency models stating that the upper kHz QPO peak is formed by
Keplerian motion at a preferred radius in the disk (e.g. the sonic
point radius, Miller, Lamb, \& Psaltis 1998), whereas the lower kHz
QPO peak formed at the frequency of the beat between the neutron star
spin frequency and this Keplerian frequency, cannot in their original
form explain the decrease in peak separation in these two sources. A
relatively small extension of the model (Lamb, Miller, \& Psaltis
1998) can, however, produce the observed decrease in peak
separation.\par
\noindent
Osherovich \& Titarchuk (1999) developed a model in which the kHz QPOs
arise due to radial oscillations of blobs of accreting material at the
magnetospheric boundary. The lower kHz QPO frequency is in their
model identified with the Keplerian frequency. Besides this QPO
two eigenmodes are identified whose frequencies coincide with the
upper kHz QPO peak frequency and the frequency of the HBO component in
the power spectra of Sco~X--1 (Titarchuk \& Osherovich
1999). Interpreting our findings within this framework did not result in
stringent constraints on the model.  \par
\noindent
We found that the peak separation is consistent with being constant
(Fig.~\ref{kHz_vs_HBO_two_selection} A and B), but neither a decrease
towards higher $\dot{M}$ as in Sco~X--1, 4U~1608--52, 4U~1735--44,
4U~1702--429, and 4U~1728--34 nor a decrease towards lower $\dot{M}$,
as predicted by Stella \& Vietri (1999) can be ruled out. If the model
of Stella \& Vietri turns out to be the right one the mass of the
neutron star most likely is in the range of 1.8 to 2.2 $M_{\odot}$
(see Fig.~\ref{kHz_vs_HBO_two_selection} B). This is in agreement with
the mass of Cyg~X--2 derived by Orosz \& Kuulkers (1999), and with the
masses of the neutron stars derived when interpreting the highest
observed kHz QPO frequencies as due to motion at or near the
marginally stable orbit (Kaaret, Ford, \& Chen 1997; Zhang,
Strohmayer, \& Swank 1997).

\acknowledgments This work was supported in part by the Netherlands
Organization for Scientific Research (NWO) grant 614-51-002. This
research has made use of data obtained through the High Energy
Astrophysics Science Archive Research Center Online Service, provided
by the NASA/Goddard Space Flight Center. This work was supported by
NWO Spinoza grant 08-0 to E.P.J.van den Heuvel. MM is a fellow of the
Consejo Nacional de Investigaciones Cient\'{\i}ficas y T\'ecnicas de
la Rep\'ublica Argentina. Support for this work was provided by the
NASA through the Chandra Postdoctoral Fellowship grant number
PF9-10010 awarded by the Chandra X-ray Center, which is operated by
the Smitsonian Astrophysical Observatory for NASA under contract
NAS8-39073.

\end{document}